\documentclass[sigconf]{acmart}
\AtBeginDocument{%
  }

\usepackage{makecell}
\usepackage{xcolor}
\usepackage{multirow}
\usepackage{amsmath}
\usepackage{algorithmic}
\usepackage{stfloats}
\usepackage{dirtytalk}
\usepackage{colortbl}

\copyrightyear{2026}
\acmYear{2026}
\setcopyright{cc}
\setcctype{by}
\acmConference[CHI '26]{Proceedings of the 2026 CHI Conference on Human Factors in Computing Systems}{April 13--17, 2026}{Barcelona, Spain}
\acmBooktitle{Proceedings of the 2026 CHI Conference on Human Factors in Computing Systems (CHI '26), April 13--17, 2026, Barcelona, Spain}
\acmPrice{}
\acmDOI{10.1145/3772318.3790474}
\acmISBN{979-8-4007-2278-3/2026/04}

\begin{document}

\title{Conversing with Objects toward Fluid Human and Artificial Identities during Life Transitions}

%\author{Anonymous}
\author{Yuhui Xu}
\orcid{0000-0002-3136-2149}
\affiliation{%
  \institution{Eindhoven University of Technology}
  \city{Eindhoven}
  \country{the Netherlands}
}
\email{y.xu1@tue.nl}

\author{Minha Lee}
\orcid{0000-0002-7990-9035}
\affiliation{%
  \institution{Eindhoven University of Technology}
  \city{Eindhoven}
  \country{the Netherlands}
}
\email{m.lee@tue.nl}

\author{Stephan Wensveen}
\orcid{0000-0001-8804-5366}
\affiliation{%
  \institution{Eindhoven University of Technology}
  \city{Eindhoven}
  \country{the Netherlands}
}
\email{S.A.G.Wensveen@tue.nl}

\author{Mahla Alizadeh}
\orcid{0000-0002-5365-4695}
\affiliation{%
  \institution{University of Siegen}
  \city{Siegen}
  \country{Germany}
}
\email{fatemeh.alizadeh@uni-siegen.de}

\author{Mathias Funk}
\orcid{0000-0001-5877-2802}
\affiliation{%
  \institution{Eindhoven University of Technology}
  \city{Eindhoven}
  \country{the Netherlands}
}
\affiliation{%
  \institution{Eindhoven AI Systems Institute}
  \city{Eindhoven}
  \country{the Netherlands}
}
\email{m.funk@tue.nl}

%%
%% By default, the full list of authors will be used in the page headers. Often, this list is too long, and will overlap other information printed in the page headers. This command allows the author to define a more concise list of authors' names for this purpose.
\renewcommand{\shortauthors}{Xu et al.}

\begin{abstract}
People's identities change during life transitions, e.g., studying abroad. They bring everyday objects that embody memories and reflect their identities during such moves. To assist in these transitions, we ask how people's human identities could be supported by their objects through an artificial agent. This paper presents an exploratory research-through-design study around how people undergoing life transitions experience conversing with their everyday objects through a chatbot. Drawing on a two-week field deployment of a technology probe and interviews with 12 participants, we contribute (1) a conceptualization of \say{trans-embodiment} describing the asynchronous imagination of object and human identities on the chatbot, (2) empirical evidences of the resulting emotional and reflective experiences that supported identity processes, and (3) three types of trans-embodied object identities for designing conversational agents for human-object conversations. Our contributions sum up to triangulating human-agent-object identity as trans-embodiment in supporting life transitions.
\end{abstract}




\begin{teaserfigure}
\centering
  \includegraphics[width=1.0\textwidth]{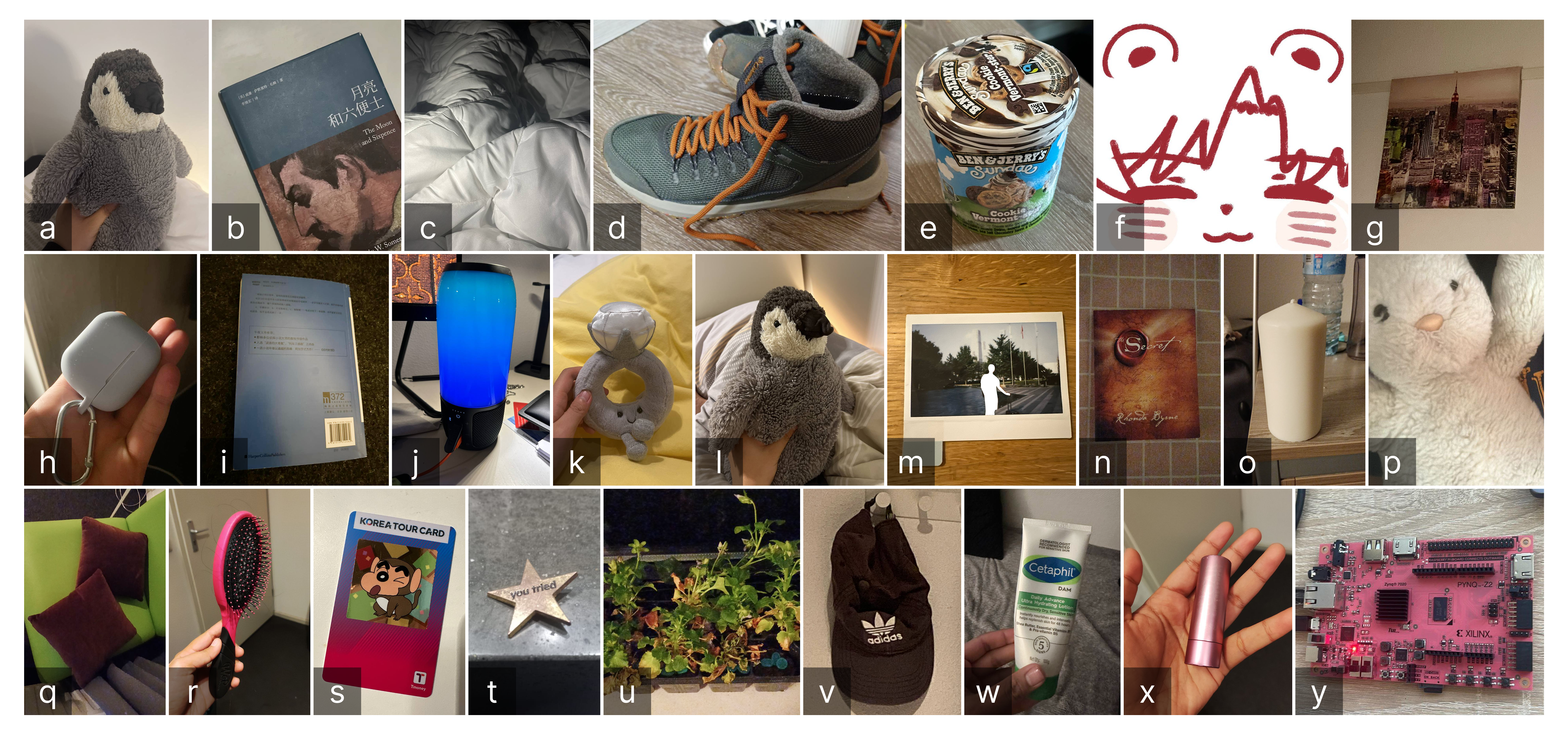}
  \caption{Participants' everyday objects (a-y) mentioned in this paper.}
    \label{objects pic}
    \Description{Twenty-five photos of participants' everyday objects mentioned in this paper from a to y in three rows.}
\end{teaserfigure}


\maketitle

\section{Introduction}

People experience life transitions when they encounter major life events, such as studying abroad. Life transitions disrupt established routines and challenge people's identities, often resulting in significant stress \cite{haslam2021life,anderson2011counseling}. To cope, individuals regulate emotions, seek social support, and engage in meaning-making and identity reconstruction activities \cite{folkman2008case,langford1997social,park2010making,mcadams2011narrative}. Among technological supports, social media is an important approach for identity construction and emotion regulation \cite{perez2024social, blumberg2016social}, which provides a space for connecting with similar others, receiving social support, and expressing or documenting transition-related experiences \cite{haimson2018social, dym2019coming, brubaker2012grief}. In recent years, Human-Computer Interaction (HCI) and Computer-Supported Collaborative Work (CSCW) communities have explored experienced identity shifts of various groups through social media platforms during life transitions \cite{haimson2019life, haimson2015disclosure, andalibi2018announcing, haimson2018relationship, gong2021smartest}. While social media significantly mediated and shaped such transitions \cite{haimson2018social}, it largely focuses on interpersonal connections and public self-expression. We therefore ask: could other technologies offer more personal, intimate ways of engaging with identity during transitions?

Everyday personal objects present such a possibility. We selectively bring everyday objects throughout our entire lives, including life transitions. They are familiar artifacts that live with us, such as cups and books \cite{mols2014making, van2015things}. They carry treasured memories, embody our social relationships, and reflect our self-identities in different periods \cite{dant1999material, belk1988possessions, csikszentmihalyi1981meaning, wheeler2021objects}. For instance, a favorite gift is cherished not only because it evokes memories with the sender, but also because its beauty reflects our aesthetic aspirations as part of our self-identities. These accumulated objects can be valuable for reviewing and supporting identities during transitions, yet they rarely \say{speak} on their own, calling for external mediation to activate their reflective and supportive potential. 

Conversational agents (CAs), such as chatbots, may offer one such mediating role. As artificial agents with natural-language interfaces \cite{dale2016return}, they can facilitate human well-being through fostering reflection \cite{kocielnik2018reflection, kocielnik2018designing, song2025exploreself} in more engaging ways than self-tracking tools or manual journaling. By asking reflective questions, CAs can prompt contemplative \cite{kahneman2003maps} and metacognitive \cite{flavell1979metacognition} thinking. HCI research has studied CAs to support well-being and reflection for various purposes, such as physical activity \cite{kocielnik2018reflection}, workspace reflection \cite{kocielnik2018designing}, and self-exploration \cite{song2025exploreself}. Meanwhile, embodied conversational agents (ECAs) as a type of CAs with physical embodiment (e.g. social robots) can also support well-being by offering emotional support and companionship with their physical presence enhancing social engagement \cite{li2015benefit, wada2007living, lee2006physically}. These studies imply that CAs might help surface implicit identity meanings of everyday objects through conversational interaction and at the same time provide companionship through objects' physical forms. 

Recent research on artificial identities further expands this possibility by revealing that artificial identities can migrate across embodiments, indicating a path for \say{direct} communication with objects. Artificial identity is an emerging area of study in human-agent interaction \cite{miranda2024robot, miranda2023examining, lee2021robo, khot2024robo, pradhan2021hey}. It refers to how agents present themselves, which can migrate across multiple embodiments or manifest differently in the same body \cite{bransky2024mind, luria2019re, chaves2018single}, e.g. Siri spanning across devices, or a single device switching between characters. This fluidity highlights how CAs' identities can be reconfigured to meet user needs and sustain extended relationships across contexts. 

\begin{figure}[t]
    \centering
    \includegraphics[width=1\linewidth]{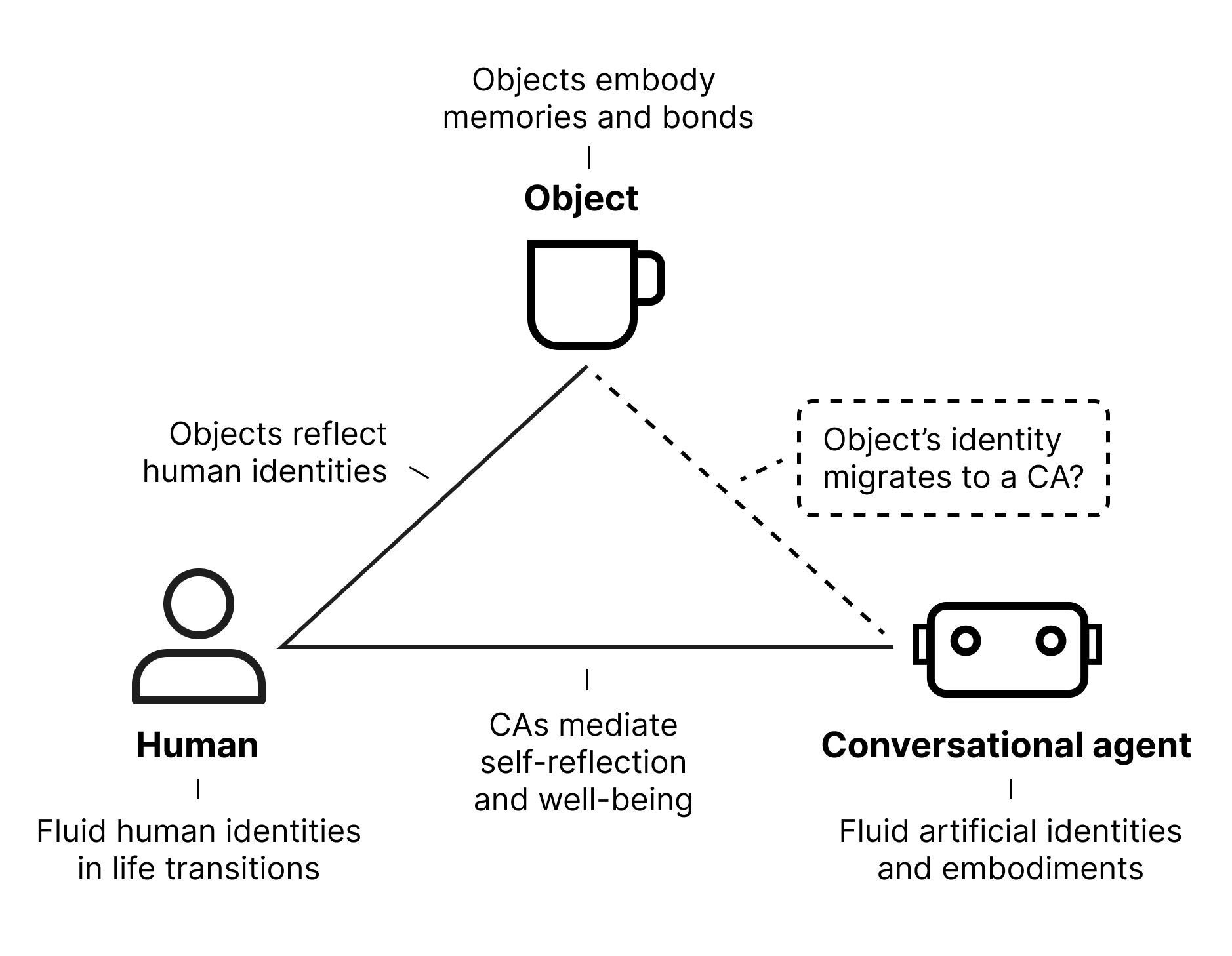}
    \caption{A triadic relationship framing among the concepts of human, conversational agent, everyday object, and their connections.}
    \label{comprehensive view}
    \Description{A diagram showing the triadic relationships among human, conversational agent, everyday object, and their connections.}
\end{figure}

In sum, we see a new opportunity to support people's identity shifts during life transitions by bringing together their everyday objects and CAs as artificial agents. Figure \ref{comprehensive view} frames this relationship as a triad of fluid human identities, fluid artificial identities, and everyday objects. It illustrates the well-established notion that everyday objects carry memories and reflect self-identities, while CAs can embody fluid artificial identities and mediate reflection. Yet, little is known about how these dynamics unfold when everyday objects and conversational agents intersect, raising new questions for exploration:

\begin{itemize}
    \item RQ1: Could a CA take on the identity of people's objects and converse with people during life transitions?
    \item RQ2: How could people perceive an object's identity through a CA? 
    \item RQ3: How could such experiences unfold in practice? 
\end{itemize}

To explore this opportunity, we conducted a research-through-design field study with 12 international students who had recently moved to the Netherlands. We designed and deployed a chatbot probe that role-played participants' personal objects for two weeks in their daily lives. We chose international students studying abroad as a case of major life transition involving cross-cultural identity reconstruction  \cite{ecochard2017international, marginson2014student}. They often bring only a few meaningful objects with them \cite{krtalic2023things} that carry subtle yet profound markers of shifting cultures, languages, and social roles \cite{leong2015coming, mori2000addressing, weiss1975loneliness}, making this context a rich lens for studying identities during life transitions, with implications for other situations such as migration or starting a new job abroad. 

Our findings introduce the concept of \textit{trans-embodiment}, i.e. the way identities of objects and object-related others can be imaginatively embodied in a chatbot across forms asynchronously. Through these trans-embodied interactions, participants experienced embodied emotional and reflective engagements that supported their transition. We further identified three types of trans-embodied object identities that emerged when the chatbot was imagined as participants' objects. Together, our findings show conversing with objects through a CA as a potential approach to support life transitions, and how enabling trans-embodiment can extend human-agent-object identity relationships in HCI. At a higher level, our work bridges HCI research on life transitions, objects, conversational agents, and artificial identity multi-embodiment, with the following three contributions:

\begin{enumerate}
    \item We provide empirical evidence and an in-depth account of how people undergoing life transitions experience conversing with objects through a chatbot, thereby extending HCI research on life transitions with a novel approach and bridging HCI object research.
    \item Based on our findings, we conceptualize \textit{trans-embodiment} as a new way of understanding artificial identities, highlighting how a chatbot can carry fluid identities of objects and object-related others. This complements existing models of artificial identity multi-embodiment.
    \item We categorize three types of trans-embodied object identities that emerged in conversations, offering design considerations for aligning chatbot identities with people's imaginations of objects through trans-embodiment.
\end{enumerate}

\section{Related Work}

In this section, we review five areas of related work that frame our study. We begin with research on life transitions in HCI as the broader context. We turn to everyday objects as resources for self-identity reflection with their rich associated memories and meanings, followed by CAs that mediate reflections and well-being through conversations. Lastly, we examine work on artificial identities and embodiments to illustrate the space where CA may embody objects to converse with people to mediate reflections on object-indicated self-identities.

\subsection{Life Transition in Psychology and HCI}

Life transition is the transitional period after experiences or events that \say{\textit{disrupt a person's established and routinized patterns of behavior}} \cite[p. 636]{haslam2021life}. Life transitions often bring stress, as individuals' identities are challenged when adapting to new roles, routines, and relationships \cite{anderson2011counseling, haslam2021life}. To cope, people commonly regulate emotions (e.g., expressive writing) \cite{folkman2008case, batenburg2014experimental, baikie2005emotional}, draw on social support \cite{langford1997social}, or engage in meaning-making processes through reappraising events \cite{skaggs2006searching,park2010making}. Identity management constitutes a specific coping mechanism for identity-relevant stress. For instance, social identities are maintained or reshaped through old and new group memberships \cite{haslam2021life}, while personal identities are often negotiated via autobiographical narrative activities that integrate past, present, and future experiences \cite{mcadams2011narrative, pasupathi2007developing}. In this way, managing one's identities both responds to and mitigates the stress arising from life transitions. 

HCI and CSCW researchers have conducted extensive studies on life transitions in relation to social media as a type of technological support \cite{dym2019coming, haimson2015disclosure, brubaker2012grief, andalibi2018announcing, semaan2016transition, moncur2016role}, such as gender transition \cite{haimson2015disclosure}. The objectives are to understand the experiences of people in various life transitions interacting with social media \cite{haimson2021major, haimson2016digital, haimson2018social, wilcox2023infrastructuring, zhang2022separate} and to explore designing technology-supported life transition experiences \cite{chuanromanee2022designing, haimson2023uncovering, regis2025embracing}. Haimson \cite{haimson2018social} proposed the concept of \say{social transition machinery} to describe how multiple, often separate, social media sites work together to support life transitions, particularly in the context of gender transition. Zhang et al. \cite[p. 458]{zhang2022separate} further examined motivations and challenges in navigating these separate networks. With implications for the design of social media, Haimson and Marathe \cite{haimson2023uncovering} showed that sentiment visualizations of social media data can trigger reflection and facilitate interpretation of identity changes, especially for marginalized individuals. However, some experiences, such as pregnancy loss \cite{andalibi2020disclosure}, might be difficult to disclose on social media due to stigma, limiting opportunities for social support. This highlights the complexity of using social media during vulnerable life transitions, where the potential for support is often constrained by concerns over disclosure. 

\subsubsection{International students and life transition}

International students entering higher education represent a typical group undergoing life transitions. They experience separation from established social networks (e.g., family and friends) \cite{weiss1975loneliness}, past and new identities in flux \cite{constantine2005qualitative, leong2015coming}, and incorporation of the identities and social networks \cite{marginson2014student, xu2018role}. Compared to local students, they face unique sources of stress, such as language difficulties and maintaining close relationships \cite{mori2000addressing}, loneliness due to loss of supportive social networks \cite{weiss1975loneliness}, and the challenge of building social connections with domestic students \cite{leong2015coming}. These situations that are difficult to understand without similar cultural experiences may prevent international students from disclosing their struggles to even close relationships \cite{constantine2005qualitative}. For adaptation, international students continuously self-reflect, during which their self-identities are reshaped \cite{marginson2014student, xu2018role}. However, international students' life transition through technology have drawn little attention from the HCI and CSCW communities \cite{binsahl2015identity, sabie2021migration}. Thus, we consider this group's life transition experiences a valuable case for generating insights that are generalizable to other identity-shifting populations, while also addressing this group's unique needs.

\subsection{Everyday Objects, Identity, and HCI}

Everyday personal objects are not only associated with memories, meanings, and future dreams \cite{csikszentmihalyi1981meaning}, but also embodiments or representations of human identities. Literature on psychology and material culture supports that everyday objects play roles in reflecting and shaping identities of ourselves and others within social networks \cite{wheeler2021objects, dant1999material, belk1988possessions}. A recent literature review by Wheeler and Bechler \cite{wheeler2021objects} applied theories from psychological consistency and proposed a five-panel model of the relationships among one's self-identity, object, and others, clarifying that objects, despite being highly symbolic, can play a central role in one's private and public identity formation and expression. Similar to migrants, international students bring personal objects when studying abroad, which also symbolize and remind their cultural identity \cite{pechurina2023scaling, krtalic2023things}. 

Earlier HCI studies explored everyday objects as memory cues for interaction design for remembering, reminiscence, and reflection \cite{tsai2018memory, zijlema2019qualitative, van2015things, zijlema2016companions, tsai2017designing, jeung2024unlocking}. Tsai and van den Hoven \cite{tsai2018memory} examined objects through the Memory Probe and revealed objects with traces as embodiments of others to help reconstruct values, emotions, and expectations. Utilizing large language models (LLMs), recent works have explored various technological approaches to enabling human-object conversation \cite{iwai2025bringing, nagano2023talk, wang2025talking} as a long-standing human desire. For example, to animate and converse with objects, Iwai and Matulic \cite{iwai2025bringing} built an LLM-powered augmented reality (AR) mobile application for text-based interactions, while Wang et al. \cite{wang2025talking} approached it through an LLM-powered wearable system enabling real-time voice interactions. However, beyond animating objects through visual appearance or functionality, these systems pay limited attention to objects' holding memories and meanings that could strongly shape the human-object relations and interactions \cite{csikszentmihalyi1981meaning}. In sum, we saw that these various domains of HCI object studies are fragmented and rarely build on each other's strengths, calling for bridging works to allow deeper engagement of human-object interactions. Moreover, how objects can indicate human self-identities is overlooked in supporting reflection, indicating a research gap in leveraging this potential of objects.

\subsection{Conversational Agents for Reflection and Well-Being}

CAs are mediating human well-being. There are HCI studies on text-based CAs, e.g. chatbots, that foster people's well-being through various approaches, such as facilitating reflection \cite{kocielnik2018reflection, kocielnik2018designing, song2025exploreself}, triggering self-compassion \cite{lee2019caring}, and regulating emotions \cite{denecke2020mental}. For instance, Song et al. \cite{song2025exploreself} developed a conversational assistant called ExploreSelf driven by an LLM, for individual reflections on personal challenges. Denecke et al. \cite{denecke2020mental} developed and tested a chatbot application called SERMO that implements cognitive behaviour therapy methods to support regulating emotions and dealing with thoughts and feelings. Karaturhan et al. \cite{karaturhan2024informing} developed a set of interaction strategies for designing reflections with CAs based on CAs' contextual awareness, statement repetition, and human-likeness. 

ECAs are another type of conversational agents with a virtual or physical embodiment that enables verbal and non-verbal communication \cite{isbister2004blind, ruttkay2004embodied}. Physical ECAs, such as social robots, are researched and used in well-being contexts for emotional support and companionship \cite{ullrich2016murphy, saldien2010expressing, li2015benefit, wada2007living, gamborino2019mood}, as their physical embodiments may enhance social presence and engagement \cite{lee2006physically, segura2012you, deng2019embodiment, oh2020differences}. This enhanced social experience of physical ECAs is further explained by their inherent materiality that mutually and importantly constitutes their social and agential qualities \cite{alavc2016social}. 

However, CAs and ECAs mediating well-being can raise ethical concerns in long-term usage, such as emotional dependence \cite{laestadius2024too, zhai2025unpacking, xie2022attachment}. Laestadius et al. \cite{laestadius2024too} found that social chatbots designed for companionship (e.g. Replika) can lead to emotional dependence, where users respond to human-like needs and emotions from the chatbot, potentially harming their mental well-being. ECAs can also raise additional ethical concerns due to their embodied forms. High levels of human-likeness can elevate users' expectations \cite{kraus2016human, van2014robot} and encourage suspension of disbelief \cite{heckman2000put}, i.e. temporarily allow oneself to believe something that is not true\footnote{\url{https://web.archive.org/web/20180729045913/https://en.oxforddictionaries.com/definition/suspend_disbelief}}. When such expectations or trust are not met, users may experience feelings of deception \cite{fuller2011examination}, frustration \cite{rapp2024people}, or discomfort \cite{mori2012uncanny}. These examples demonstrate the risk of CAs and ECAs mediating well-being, highlighting the need for careful design and ethical consideration of their conversations and roles.

\subsection{Artificial Identities and Multi-Embodiments}

Artificial identity is an emerging research domain in human-agent interaction about \say{who is the artificial agent?} \cite{miranda2024robot, miranda2023examining, lee2021robo, khot2024robo, pradhan2021hey}. In recent years, researchers have investigated artificial identity from various aspects, such as its constitution \cite{miranda2024robot}, multi-embodiment and trust \cite{bejarano2023no}, and ethics and morality being human-like \cite{miranda2023examining, strait2018robots}. One of the popular research topics is the migration of an artificial identity across multiple embodiments \cite{bransky2024mind, luria2019re, chaves2018single, laity2025robot, tejwani2020migratable}, such as Alexa on a smartphone and a smart speaker. Research on multi-embodiment highlights the need for careful identity design to maintain a consistent agent across different bodies \cite{martin2005maintaining, koay2016prototyping, sinoo2018friendship}. Martin et al. \cite{martin2005maintaining} introduced identity cues as a means to support this consistency. Their work shows that preserving an agent's identity is central to multi-embodiment, while also posing practical challenges of recognizing the same agent across embodiments. A recent scoping review of multi-embodiment research from Bransky et al. \cite{bransky2024mind} identified identity cues, identity performance, and models of social presence as three of the research challenges of multi-embodiment identity design, pinpointing the gaps for future research. For an overview, Luria et al. \cite{luria2019re} synthesized four models of how artificial identities can be embodied in artificial agents, namely \textit{one-for-one}, \textit{one-for-all}, \textit{re-embodiment}, and \textit{co-embodiment}. Chaves and Gerosa \cite{chaves2018single} examined co-embodiment, i.e. multiple social presences co-exist on one body at the same time, in an experiment of planning a trip through a simultaneous multi-chatbot conversation. Their results indicate an exploration space for other possibilities, such as social-oriented chatbot interactions and new chatbot identity-body models. In the present study, we contribute to this area by proposing the \say{trans-embodiment} model to be introduced in Section \ref{findings}.

\section{Methodology}

This study is the first step of a multi-year research-through-design (RtD) project on technology-mediated conversation with objects. On a methodological level, we adopted RtD as an approach to generate knowledge through the making and use of prototypes \cite{stappers201443}. Method-wise, we designed and deployed a technology probe as a simple, flexible and adaptable technology for exploring user behavior, field-testing, and inspiring designs of new technology \cite{hutchinson2003technology}. In RtD, prototypes \say{\textit{play a central role in the knowledge-generating process}} \cite[Section 41.1.4]{stappers201443}, and technology probes exemplify this use of prototypes as a means of inquiry \cite[p. 267]{wensveen2014prototypes}, which is, in our case, an inquiry into the dynamic of human-agent-object identities during life transitions. Similar RtD works in HCI have used technology probes as prototypes to collect data and investigate human participants' living experiences with them \cite{odom2019investigating, chen2023exploring, kun2024exploring, li2024distancebit}.

The collected insightful data from our probe and interviews were analyzed under an abductive approach, i.e. bringing together data-driven induction and theory-guided deduction to form an explanatory theory \cite{peirce1974collected, timmermans2012theory}. For a year, we formed a \say{community of inquiry} \cite{timmermans2012theory} as a group of researchers in designing and building the probe, and understanding and discussing the data in line with related works.

\begin{figure*}[t]
    \centering
    \includegraphics[width=1\linewidth]{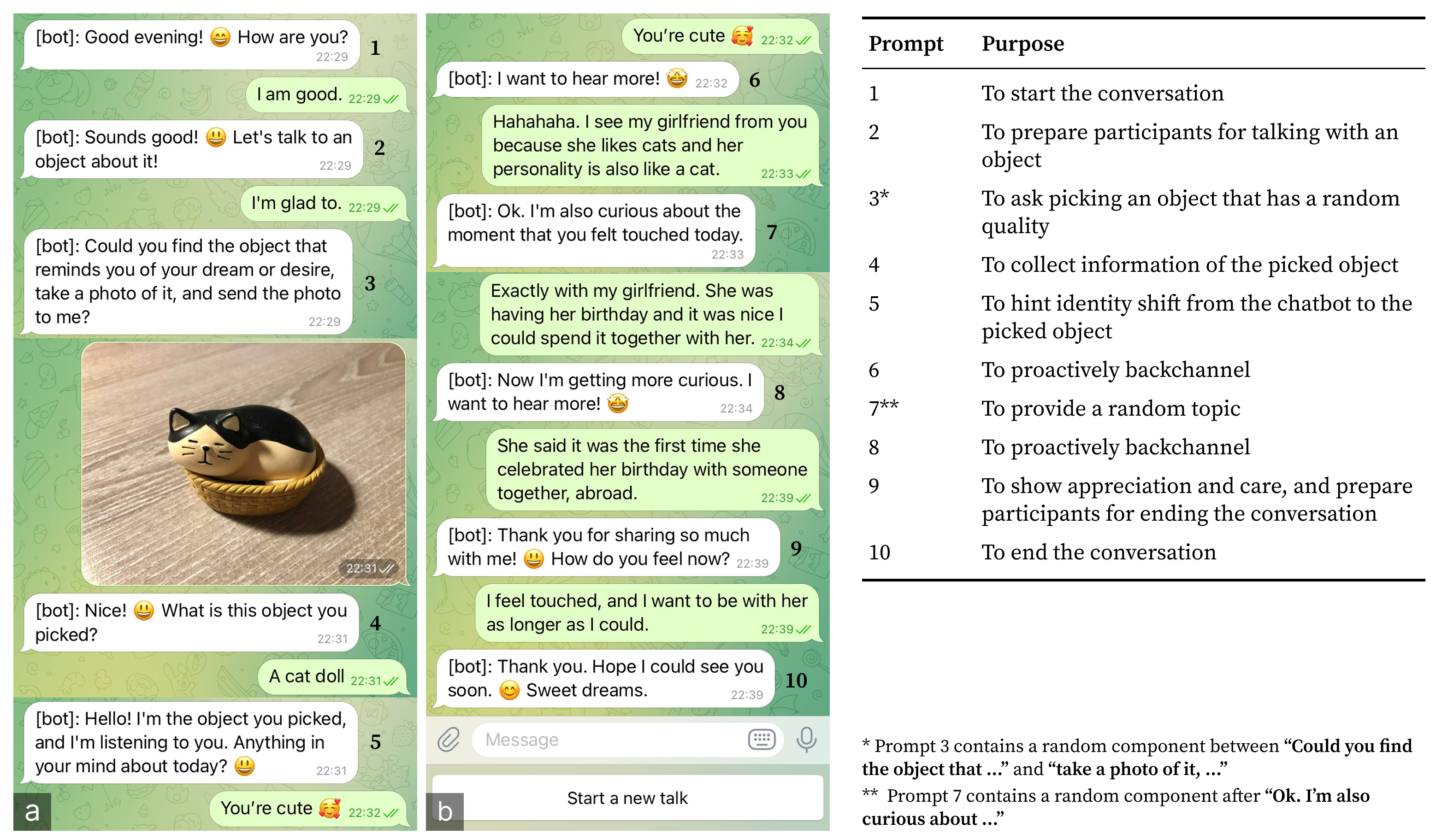}
    \caption{Left: An example of a conversation from (a) to (b) in sequence, with chatbot prompts from 1 to 10 that form the structure of the conversation. Right: Explanations of the purpose of each prompt. Footnotes: Prompt 3 and 7 each contain a random component.}
    \label{chat interface}
    \Description{Left: A chat interface of an example of a conversation with a small cat through the chatbot. From left to middle in sequence, the long interface is cut into two pieces from (a) to (b). (a): They started the talk and greeted each other, picked an object, and uploaded a photo of it, the chatbot pretended to be the object. (b): The participant started sharing, talked about another topic, and ended the talk. Right: A table of explanations of the purpose of each chatbot message, with two footnotes noting that Prompt 3 and 7 each contain a random component.}
\end{figure*}

\subsection{Probe Design Process, Rationale, and Implementation}

The technology probe is a combination of a smartphone-based chatbot (from researchers, Figure \ref{chat interface}) and varied personal objects (from participants). It is designed to examine whether and how participants can perceive and converse with their objects through a chatbot, and the experiences unfold in a life-transition background. Participants directly interacted with the chatbot as a mediator to engage with their objects. The chatbot first prompted participants to select an object, and then adopted the object's identity by stating \say{Hello! I'm the object you picked}, referring to itself as the object (\say{I}) to converse with participants. This process operationalized a \say{chatbot-as-object} interaction, giving the selected objects a voice through the chatbot. Throughout the interactions, various selected objects prompted participants differently with their associated memories \cite{van2015things}. Together, the chatbot and objects allowed participants to build their own personalized probes to probe themselves. 

Our design process began by exploring picking an object. As an RtD field study \cite{koskinen2011design}, instead of being specified by us researchers, the objects should be real from participants due to their associated memories, meanings, and self-identities \say{\textit{in natural settings}} [p. 69]. Those associations may influence how participants perceive and converse with their objects through the chatbot, and allow for a diverse range of objects to generate varied experiences. Free object selection was tested initially, but the selections were mainly driven by objects' proximity rather than meaning. We interpreted this as a lack of conscious awareness of objects' meanings in mundane everyday life \cite{miller1997material}, and thus added object-picking prompts (Prompt 3 in Figure \ref{chat interface}) to help participants attend to their surroundings and select objects purposefully. This decision was influenced by prior HCI works that applied prompts about objects for reminiscence \cite{zijlema2019qualitative, tsai2018memory}.  

The design of object-picking prompts (Prompt 3) started with specifying the content. Initially, we brainstormed prompts targeting specific objects (e.g., \say{a cup that makes you happy}), but later adopted open-ended prompts (e.g., \say{an object that makes you happy}) to respect participants' personal interpretations and preferences, inspired by \cite{tsai2018memory}. We brought up over fifty open-ended prompts varied from five aspects of objects' meanings adapted from \cite{csikszentmihalyi1981meaning}, i.e. daily activity, emotion, memory, symbolic value, and relationship. Meanwhile, we explored using OpenAI GPT-3 models, such as Babbage\footnote{Babbage-002 model: https://platform.openai.com/docs/models/babbage-002}, to generate varied prompts based on the five aspects in real time during the interactions, but the outputs were unstable in content and format. In the end, we stayed with using polished brainstormed object-picking prompts (Prompt 3) as presets and reduced the number to 21 to be more manageable. In conversations, we allowed participants to choose an object freely and naturally after getting an object-picking prompt (Prompt 3), to motivate them in scanning and interpreting objects around, and to capture a variety of objects with different meanings. 

After picking an object, we asked participants to upload a photo and input its name to capture information of its appearance and contextual features. This allowed us to better understand and empathize with participants' thoughts and emotions in conversations, informed by the cultural probe methodology \cite{gaver1999design}. The identity shift prompt (Prompt 5 in Figure \ref{chat interface}) functioned as a conceptual identity shift signal \cite{laity2025robot} from the chatbot to the selected object. Instead of generating explicit identity cues through AI models, we intentionally kept the chatbot's identity vague and neutral, allowing participants' free interpretations based on their stories with the objects rather than adopting predefined ones. 

When designing the conversation after the identity shift prompt (Prompt 5 in Figure \ref{chat interface}), we aimed to let the chatbot show a non-judgmental attitude to encourage disclosure rather than provide advice. We scripted the responses and did not apply AI models for personalized ones due to data privacy concerns. Initially, the responses to \say{Anything in your mind about today?} were short and less motivated, thus we added randomized proactive backchanneling prompts (Prompt 6 and 8 in Figure \ref{chat interface}) to encourage engagement, inspired by prior HCI work \cite{ding2022talktive}. We further designed topic prompts (Prompt 7 in Figure \ref{chat interface}) to provide a topic in a scripted conversation. Similar to how we generated object-picking prompts (Prompt 3), we also tried the AI models but stayed with the brainstormed topics prompts (Prompt 7) on positive and neutral experiences on that day. This aimed to keep conversations open to present experiences, to enable self-event connections and narratives through objects \cite{pasupathi2007developing}, which also applied to asking about the day. We designed both the object-picking prompts (Prompt 3) and topic prompts (Prompt 7) to be either positive or neutral to prevent negative experiences, such as rumination \cite{takano2009self}. They were both randomly drawn each time from two separate component pools that were the same for all participants. The randomization aimed to offer a positive experience and stimulate interpretation \cite{leong2006randomness}. The conversation after the identity shift prompt (Prompt 5) was kept in five rounds before the end to prevent boredom and burden. 

We implemented the chatbot based on an internally developed Telegram bot\footnote{Telegram Bots: https://core.telegram.org/bots}, i.e. a small application that runs entirely within the Telegram app. Our department maintains an internal platform called Data Foundry\footnote{Data Foundry: https://data.id.tue.nl/documentation/about} that provides a fully configured Telegram bot infrastructure for collecting in-the-moment data (e.g., diaries, photos). This setup provided a secure data environment, as the bot chat messages were sent directly to our own server.\footnote{Terms of Service for Telegram Bots: https://telegram.org/tos/bots} The internal platform also includes a JavaScript-based integrated development environment (IDE) that allows us to script the chatbot's conversation flow and data handling, and provides APIs for LLM abilities. One limitation of our implementation is that the Telegram bot chat did not support displaying a bot avatar, preventing us from investigating identity shifts conveyed through visual cues. Yet, this internally developed Telegram bot offers the most accessible, robust, secure, and cost-effective solution. 

\subsection{Participants and Data Collection Process}

We recruited \textit{international students who had recently moved to the Netherlands} as participants. We posted ads in local migrant-specific groups on social media and connected to international students in Eindhoven through word-of-mouth. People who expressed interest in our study were invited to a brief introductory session. Twelve participants joined our study (Table \ref{participants table}). All of them arrived in the Netherlands within one year and were studying at the same university. 

\begin{table}[t]
\caption{Demographics of the 12 participants.}
\label{participants table}
\centering
\begin{tabular}{l>{\centering\arraybackslash}p{0.25\linewidth}>{\centering\arraybackslash}p{0.25\linewidth}>{\centering\arraybackslash}p{0.25\linewidth}} 
\toprule
 & \textbf{Age } & \textbf{Gender } & \textbf{Nationality } \\ 
\midrule
P1& 25& Man& Chinese\\ 
P2& 23& Woman& Chinese\\ 
P3& 21 & Man& Hungarian \\ 
P4& 21 & Woman& Hungarian \\ 
P5& 23 & Woman& Chinese \\ 
P6& 25 & Woman& Chinese \\ 
P7& 20 & Woman& Indonesian \\ 
P8& 24 & Woman& Chinese \\ 
P9& 23 & Man& Italian \\ 
P10& 24 & Woman& Indian \\ 
P11& 33 & Man& Iranian \\ 
P12& 23 & Woman& Indian \\ 
\bottomrule
\end{tabular}
\end{table}

The chatbot prompted participants every night at 8 pm between October and December in 2023. We chose this time for prompting because it allows people to reflect on the experiences accumulated throughout the day. The data collection lasted for two weeks to collect long-term and experiential data in various contexts and to avoid potential data distortion caused by the novelty effect \cite{gravetter2009research}. Every night at 8 p.m., a notification was sent to the participants as a reminder to do the task before bed. Within a week after the two-week probe data collection, we invited the participants individually to a one-hour semi-structured interview to review their conversations and the motivations behind their words. Eventually, we collected 146 conversations and 713 minutes of interview data. All participants completed the data collection without dropping out in the middle. Each participant received a 20 euros gift card after the study. 

\subsection{Ethical Considerations } \label{ethical considerations}

This study was approved by the Ethical Review Board (ERB) at Eindhoven University of Technology. Psychological and social support resources were available in campus, including student psychologists, student counselors, academic advisors, confidential peer supporters and advisors, and more (i.e. eleven support offices in total\footnote{https://educationguide.tue.nl/guidance-and-development/who-to-contact}) to support students who could be negatively affected. Participants were asked about their current mental health status. All reported no severe psychological concerns. In an in-person kick-off meeting, participants were informed of the data collection procedures and potential risks, reminded of their absolute right to withdraw at any time without explanation, and provided informed consent for the anonymized use of their data in research publications. During data collection, we monitored the conversations and reviewed the sensitive and vulnerable data within the team. After the first week, we conducted a halfway meeting with each participant to check on their experiences and well-being. A similar review was conducted during the one-on-one exit interview. After data collection, participants' access to the chatbot was immediately terminated. These measures aimed to reduce any potential negative impact on participants' well-being.

\begin{figure*}
    \centering
    \includegraphics[width=1\linewidth]{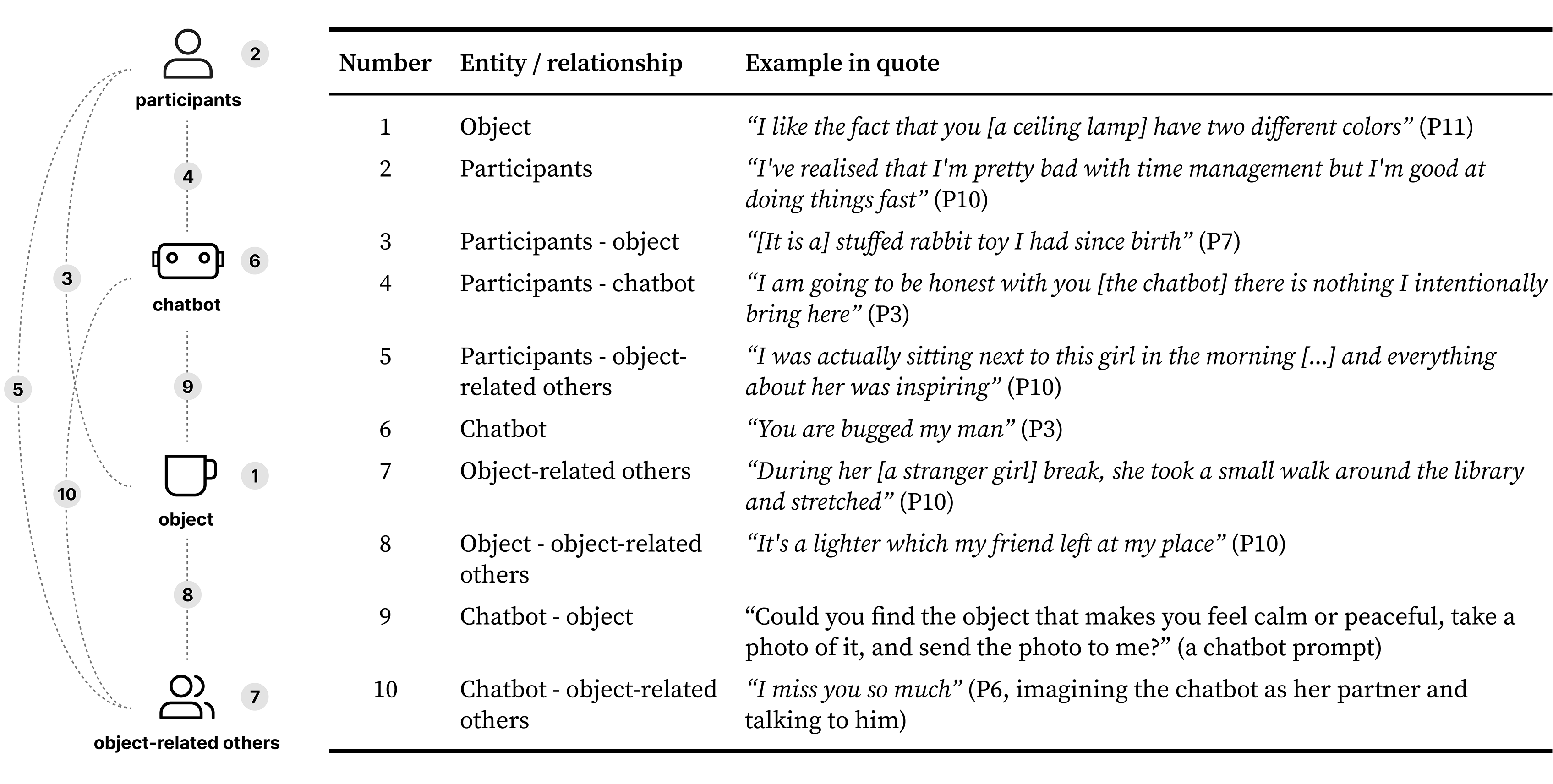}
    \caption{Left: An entity-relationship diagram that describes the parties involved and their interrelationships in this study from 1 to 10. Right: A table that explains the four entities and six interrelationships from 1 to 10 with examples in quotes.} 
    \label{ER diagram}
    \Description{Left: A diagram with 4 entities in icons inter-connected with 6 lines, all indicated with numbers from 1 to 10. The entities are participants, chatbot, object, and object-related others. Right: A table with three columns: "number", "entity / relationship", and "example in quote" that explain the four entities and 6 relationships on the left diagram. }
\end{figure*}

\subsection{Data Analysis}

We used abductive thematic analysis \cite{braun2006using, timmermans2022data, thompson2022guide} to analyze the data. Thematic analysis allows identifying patterns \cite{braun2006using}, which aligns with our abductive approach of generating novel interpretations from surprising observations \cite{timmermans2012theory}. We dealt with participants' demographic data, interview data, texts to the chatbot, and images taken by participants. To assist, we made a diagram (Figure \ref{ER diagram} left) to represent all the entities and relationships involved in the dataset, and defined the criteria (Figure \ref{ER diagram} right) for data to match each entity or relationship, followed with quotes. Through the diagram, we were able to capture both participants' situated interaction experiences (participants - chatbot - object) and their broader ongoing life transition experiences (participants - object - other parties) in the dataset. 

The abductive thematic analysis was guided by works from Timmermans and Tavory \cite{timmermans2022data} and Thompson \cite{thompson2022guide}. Firstly, in the open coding phase, the first author coded the dataset in MAXQDA and discussed initial codes with the research team, iteratively grouping them and identifying \textit{relationships} as a possible organizing analytical theme. Secondly, in the focused coding phase, we dived into \textit{how these relationships potentially influenced the conversations}, coding relevant data to deepen this aspect and building patterns and themes across entities. Finally, in the theorizing phase, we revisited and refined codes and sub-themes into higher-level themes, which were iteratively reviewed and discussed until the final thematic structure was established as presented in the findings.

\subsection{Positionality}

We describe our positionality as it may influence our data analysis. Collectively, our team brings interdisciplinary expertise in software engineering, HCI, ethics, design, and electrical engineering, as well as diverse cultural backgrounds spanning East Asia, the Middle East, and Western Europe. We all share experiences of being international students or living abroad, which resonated with participants' identity tensions during the life transition of studying abroad. We acknowledge that this shared experience might have sensitized us to certain data (e.g., acculturation, identity shift) while potentially making other interpretations less salient. Although we did not formalize procedures for emotional reflection, we remained empathetic, respectful, and careful in our analytic stance. Throughout the analysis, we engaged in regular discussions to surface assumptions, question our interpretations, and consider alternative readings of the data. This reflexive process helped us remain aware of how our positionality shaped, instead of dominated, our analytic decisions.

\section{Findings} \label{findings}

This study explores \say{conversations with objects through a CA} as a technological approach to support identity shifts during life transitions, giving objects a voice through leveraging artificial identities' multi-embodiment. Despite using qualitative methods, we first present descriptive statistics of the conversation data to provide an overview, given that our analysis focuses on how the relationships potentially influence the conversations. The number of all valid conversation entries is 135. There were 82 (60.7\%) conversations observed from all 12 participants as \say{influenced} by the human-object relationships. Meanwhile, 39 (28.9\%) conversations were identified as having \say{no influence} from the human-object relationships, and 14 (10.4\%) conversations were identified as not strongly influenced due to the lack of relative information. 

Next, we present the findings of the abductive thematic analysis. We begin with a brief background identified from the data to demonstrate participants' vulnerable transition experiences and to contextualize the themes. The themes then reveal, when interacting with the probe within the two-week study, how participants perceived the chatbot's identity shifting through their objects and object-related others (RQ1), and then how they came to see their objects through the chatbot in three ways (RQ2). These identity shifts then brought participants embodied emotional and reflective experiences (RQ3). 

Moving abroad for study was a hard transitional period. Our participants were aware of their encounters with various stressors and loneliness. They felt stress from their studies (all), language difficulties (P8, P9), livelihood (P5, P7, P10, P12), or dissatisfaction with self (P4, P6, P7, P9, P12). These aspects formed accumulated stress and negative emotions. For example, P8 once said to her penguin doll (Figure \ref{objects pic}a), \say{\textit{I feel that studying abroad is very painful and I want to return to my country}}. What is worse, they could not get enough quality support from the people around or far away. Meanwhile, establishing new relationships in the new place was stressful, such as collaborating with teammates (P1, P5, P8), making new connections with locals (P8, P9, P12), and maintaining new friendships (P1, P9, P12). P5 shared with her book (Figure \ref{objects pic}b) about her bad experience of \say{\textit{being squeezed out by [her] local classmates}} and \say{\textit{even some discrimination}}. Sometimes they also faced the challenge of maintaining intimate relationships with partners (P3, P6, P10) or friends (P10, P12) far away. P6 once had an argument with her boyfriend, which \say{\textit{almost led to [them] a breaking up}}. These challenges could add to their loneliness and missing their close people, such as family (P1, P3, P7, P10), partners (P4, P5, P6, P8, P9), friends (P2, P5, P10), and important persons such as an idol (P5). For instance, P5 shared her implicit longing for her partner toward her thick quilt (Figure \ref{objects pic}c) and said, \say{\textit{my lover told me that it was going to first snow where he was. It's a pity that I can't see it with him}}. These vulnerable experiences were constantly impacting and shaping our participants' social, cultural, and self-identities throughout the transitional period.

\subsection{Trans-Embodiment: Fluid Identities of Human, Agent, and Object} \label{trans-embodiment-section}

\begin{figure*}
    \centering
    \includegraphics[width=0.85\linewidth]{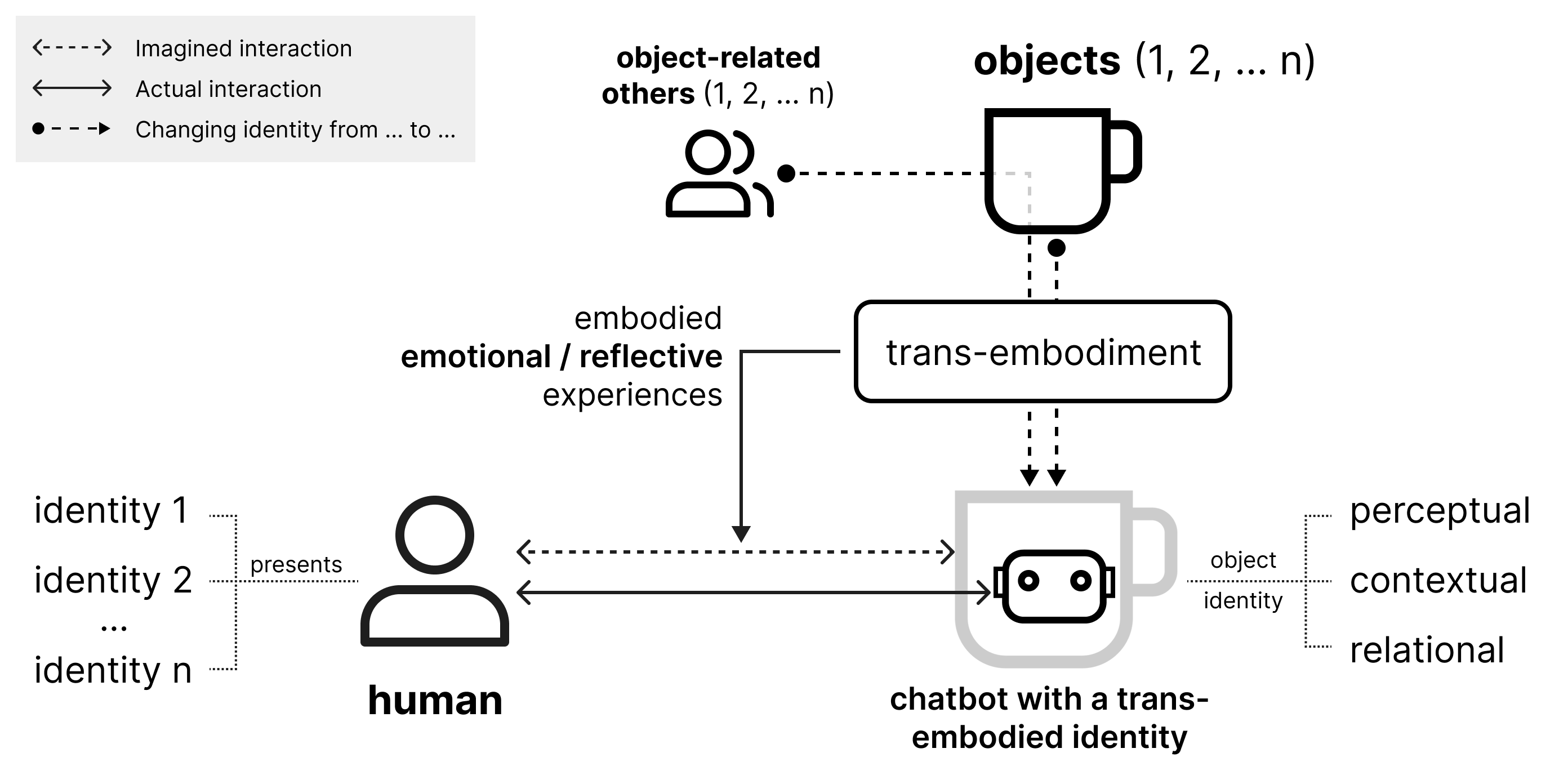}
    \caption{Trans-embodiment: a presence in which the identities of objects (e.g., cups) and object-related other humans in participants' lives can be imagined on the chatbot across forms asynchronously. These trans-embodied identities are based on perceptual, contextual, and relational imaginations of objects. Meanwhile, people present different social identities correspondingly, resulting in embodied emotional and reflective experiences through the conversations. }
    \label{trans-embodiment}
    \Description{A diagram of trans-embodiment. From left to right, human identities are connected to human, human is interacting with a chatbot but imagining interacting with a cup. The chatbot is representing a cup as an object, and the personal objects are either with their own identities or embodying another person's identity.}
\end{figure*}

Here we answer RQ1: \textit{Could a CA take on the identity of people's objects and converse with people during life transitions?} We found that when conversing with the chatbot, participants not only projected object identities onto it but also attributed other human identities. Hence, the boundaries among the identities of artificial agents, humans, and objects were blurry. To capture this phenomenon, we introduce the concept of trans-embodiment in Figure \ref{trans-embodiment}, i.e. a presence in which the identities of objects and other humans in participants' lives can be imagined on the chatbot across forms asynchronously. Trans-embodiment shaped how participants presented their own identities in conversation. As the chatbot took on the identities of different objects or humans, participants adjusted both the content and manner of their conversation, which in turn brought forth emotional and reflective experiences. Our themes are summarized in Table \ref{table:themes}. 

\begin{table*}[t]
\renewcommand{\arraystretch}{1}
\caption{Themes and sub-themes in the present study.} 
\label{table:themes} 
\resizebox{\textwidth}{!}{
\begin{tabular}{lp{7cm}p{7cm}}
\toprule
\textbf{Trans-embodiment} & \textbf{Human, artificial, and object identities can be trans-embodied as emotional and reflective experiences} \\ 
\midrule
\textbf{Fluid Identities of Human, Agent, and Object}  & \textit{\textbf{Changing identity through objects}}: People could imagine the identity of their picked objects on the artificial agent.  
\\ & \textit{\textbf{Changing identity through object-related others}}: People could imagine the identities of object-related others on the artificial agent.
\\ & \textit{\textbf{Objects' identities in conversations}}: An object can be imagined on the chatbot with a perceptual, contextual, or relational identity.
\\ \midrule
\textbf{Embodied Emotional Experience} &  \textit{\textbf{Expressing stress to companion objects}}: People expressed stress to their personal objects as companions.
\\ & \textit{\textbf{Longing and support through objects' embodied others}}: People expressed longing to or got support from their other persons that were imagined onto the objects.
\\ & \textit{\textbf{Feeling safe and familiar to externalize emotions}}: Trans-embodied identities brought people a sense of safety and familiarity to externalize emotions. 
% \\ & \textit{Bringing sense of familiarity}: The embodied meanings and memories on objects bring sense of familiarity. 
\\ \midrule
\textbf{Embodied Reflective Experience}  & \textit{\textbf{Rediscovering the self from identified objects}}: People rediscovered things about themselves from their personal objects that they identified with. 
\\ & \textit{\textbf{Reviewing relationships with objects' embodied others}}: People reviewed their relationships with other persons in their lives embodied by the objects. 
\\ & \textit{\textbf{Revisiting deep bonds with beloved objects}}: People revisited their bonds with their personal objects. 
\\ 
\bottomrule
\end{tabular}}
\end{table*}

\subsubsection{Changing identities through objects}

Experiencing separations and being unintentionally isolated led to a lack of quality and supportive social connections. When provided with a chance to converse with familiar everyday objects, participants imagined the identity of their selected objects on the chatbot to fulfill their need for social connections. This moment typically occurred when the chatbot explicitly stated \say{Hello! I'm the object you picked}, before which participants (P1, P2, P3, P7, P8, P11) were still talking \textit{about} the object to the chatbot. We interpret this shift as participants began to project the object's identity onto the chatbot once it was imagined to have a perceptual or contextual identity. In doing so, participants also presented a different facet of their own identity to engage in the conversation. This change altered the content and style of the conversation, enabling participants to disclose personal matters related to the object under another self-identity mask. For example, P3 selected a pair of boots (Figure \ref{objects pic}d) as the object that \say{brings a lot of joy or happiness} because he was highly satisfied with them and believed they made him look good when wearing them. In the conversation, he first responded to the chatbot's question \say{What is this object you picked?} by only saying \say{\textit{It is a boot}}, while immediately switched to a cheerful and excited tone to talk to the \say{boots}, sharing a travel experience that also brought him joy and happiness: \say{\textit{Sooo today we traveled far, don't we? We are in Paris, baby! I am so happy to be here [...] Tomorrow, after we have an amazing breakfast, we will go to some extraordinary places. I am sooo excited about all of them. I hope you are too, boot.}}

\subsubsection{Changing identities through object-related others}

Quality support from established close relationships was limited by time and space when studying abroad. The desire to talk to significant others and be accompanied motivated participants to imagine their identities on the chatbot through some objects that were strongly bonded with them (P1, P6, P8, P11). Participants then engaged with the chatbot in a tone resembling their real-life relationship with that person. We interpret this as showing that the relational identities of objects shaped not only how participants perceived the chatbot but also how they expressed themselves. By engaging with the chatbot as if it were the associated person, participants altered the conversation's content and style, enabling them to express feelings toward the \say{person} without directly disclosing them in real life. For example, P1 often expressed life stress and anxiety to other objects during the study, such as describing a quarrel with a friend to a bowl of ice-cream (Figure \ref{objects pic}e) and saying \say{\textit{Shit. Sometimes I feel that I'm the only person that tried to keep calm and polite to others}}. When talking to a photo of her loved cartoon character (Figure \ref{objects pic}f) as her \say{\textit{spiritual support}}, she put on a different mask of a calm, independent, and resilient identity, as if not to let it worry, \say{\textit{My life here is good. I don't feel lonely or depressed anymore [...] Sometimes I would think about your words [...] Just be brave. I think it is what you want to say. Yes, I do feel more brave thinking about you. Nothing will beat me down.}} People could also talk with a \textit{place} via the chatbot. P11 imagined the chatbot as New York City in a painting (Figure \ref{objects pic}g) as the picked object and said \say{\textit{You are the city I wish to visit}}. This indicated the fluid transition of non-human identities to the chatbot, encompassing objects as places. 

\subsubsection{Trans-embodied object identities} \label{objects identities}

Here we answer RQ2: \textit{How could people perceive an object's identity through a CA?} We found that our participants imagined each object as having an identity(s) in conversations through the chatbot based on participants' bonds, knowledge, and memories of the object. We then categorized objects' trans-embodied identities into three types: \textit{perceptual}, \textit{contextual}, or \textit{relational}. Note that these three types were not mutually exclusive. 
\begin{enumerate}
    \item A \textit{perceptual} trans-embodied object identity was imagined and interpreted based on sensory traits (e.g., visual, tactile, auditory) of the object, such as \say{\textit{warm and fluffy}}. Such objects are often bonded only with the participants, as the bonds with other real persons would quickly pull the object's imagined identity toward those persons. For example, P11 frequently used his AirPods (Figure \ref{objects pic}h) only by himself in everyday life, and he considered it \say{\textit{cute}} based on its auditory trait of \say{\textit{making cute noise from closing the lid}}.
    \item A \textit{contextual} trans-embodied object identity was imagined based on experiences of the contexts where they used to interact with the object, such as working, entertaining, or socializing. This type of object embodies the experience in a certain context that may include non-specific others. For instance, in P2's case, talking to his book (Figure \ref{objects pic}i) made him think of \say{\textit{work or office}} and his speaker (Figure \ref{objects pic}j) of \say{\textit{calm and relaxed like chatting with a close friend}}. 
    \item A \textit{relational} trans-embodied object identity was imagined based on a real human-object relationship or the object's strongly associated or symbolized real person(s) in life, such as a friend, a family member, or an important person. This type of object has profound shared memories with the owner or a related person, or carries identity cues of the person, such as a photo or a gifted object pointing towards the person(s). For example, P8 was gifted a doll in the shape of a diamond ring (Figure \ref{objects pic}k) by her boyfriend, and she imagined it as her boyfriend in the conversation because it was a \say{\textit{promise of future together}} from him. 
\end{enumerate}
These imaginations of objects were later trans-embodied into the chatbot as its artificial identities. Meanwhile, the chatbot's object-picking prompts also provided a specific lens for reviewing and filtering objects. 

\subsection{Embodied Emotional Experiences} \label{Embodied Emotional Experience}

Here we answer RQ3: \textit{How could the experience of conversing with objects through a chatbot unfold in practice?} We identified embodied emotional experiences across participants and their objects. The identities of intimate objects and object-related others were trans-embodied in the chatbot, providing emotional support and enabling safe self-disclosure of vulnerable moments. In this subsection, we present three embodied emotional experiences in detail. 

\subsubsection{Expressing stress to companion objects} \label{Expressing stress to companion objects}

Some objects were attributed an identity through their qualities. These objects were usually only bonded with the participant as their owner. After moving to a new place and being separated from familiar people, participants often lacked emotional support from loved ones. In response, they turned to emotionally supportive objects for comfort, making them companions in daily life. When talking to these object companions, participants adapted their human identity to align with their role of being accompanied by the objects, resulting in conversational behaviors like expressing stress and seeking emotional support. For example, P8, who took her two penguin dolls (Figure \ref{objects pic}a and \ref{objects pic}l) to a new place intentionally for some company, vented emotions to them about feeling anxious about the collaboration with new teammates, while eventually turned to comfort herself on behalf of the doll, \say{\textit{Actually this is a temporary meeting, hmm, maybe I should relax and, and try to adjust myself}}. 

However, compared with the gradually built new relationships with other humans, the social support that objects could offer was limited and temporary, just as P12's experience described in the interview, \say{\textit{In the beginning I was responding really well. I found myself being really interested and I was giving a lot of details. Then in the in the middle I was sort of not as motivated to do it because I feel like I already have so many inner monologues and outer monologues with my friends about everything that I don't need to again have a self-reflection moment at night like I do it through the day already}}.

\subsubsection{Longing and support through objects' embodied others}

Objects with a relational identity were often bonded with other persons in the lives of the participants, such as friends, family members, or partners. When conversing with such objects, individuals tended to engage directly with the identity of the associated person and behave in line with their interrelationship. For our international student participants, these emotionally supportive individuals were often absent and far away, but the objects were physically and socially (via the chatbot) present. They expressed longing and sought comfort and support when conversing with the objects that represented those people. For example, P6 selected a photo of her boyfriend (Figure \ref{objects pic}m) as an object that \say{reminds her of a dream or desire}. Being in a long-distance relationship, the photo of her boyfriend taken during their last farewell symbolized her longing for him and her anticipation of their next meeting. In the conversation, she expressed her desire to see her boyfriend again through the chatbot, \say{\textit{I miss you so much [...] Hope to see you soon.}} In this case, her boyfriend’s identity was trans-embodied in the chatbot through the photo, allowing her disclosure of true longing as one of her identities as a girlfriend in a romantic relationship. 

Sometimes, they also responded to themselves on behalf of the trans-embodied persons. For example, P10 talked to a book (Figure \ref{objects pic}n) gifted by her boyfriend who left an encouraging message for her in it, \say{\textit{No matter where you go and do, I know that eventually you will achieve your dreams, and whenever I am at that time, I'll still be cheering for you.}} In the later conversation, she encouraged herself with what she did, influenced by her boyfriend's encouraging words, \say{\textit{I feel like maybe I need to love myself and appreciate myself more instead of branding me as stupid or useless. I guess when I feel good about myself I achieve better results.}}

\subsubsection{Feeling safe and familiar to externalize emotions}

Participants' complex emotions can be externalized safely and privately by talking to the trans-embodied identities of objects through the chatbot, because 1) the objects were physically present, 2) the chatbot was non-judgmental and could not reply, and 3) the conversations were kept private from any others. For example, P1 once had a quarrel with her roommate, and she felt \say{\textit{very wronged}}. In the interview, she revealed the difficulty of sharing it with others, \say{\textit{this matter is actually not very easy for me to talk to my parents [...] I usually don't talk to them about this, but I can't talk to others as well}}. When she had conversations with the chatbot, she felt \say{\textit{secure}} with it as a \say{\textit{safe box}} to talk to, because it was not a \say{\textit{real person}}. P8 used to post on social media, with a \say{\textit{contradictory}} hope for \say{\textit{friends to see it}} but better not \say{\textit{too many friends}}. She didn't want to \say{\textit{expose}} her \say{\textit{self-doubting words}} to other humans, but she needed inclusion from close and safe relationships. She felt \say{\textit{very safe}} to \say{\textit{vent frustrations}} to her penguin dolls (Figure \ref{objects pic}a and \ref{objects pic}l) as a \say{\textit{spiritual support}} brought from her hometown, knowing they would \say{\textit{never betray}} her. 

Also, the sense of safety came from the deep bonds, familiarity, and trust with the chatbot's trans-embodied identities from the objects and object-related others. For example, P3 once talked to his candle (Figure \ref{objects pic}o) about feeling bad about the local education system. In the interview, P3 expressed his feeling safe to talk about it to his objects as \say{\textit{these objects are like my friends, they know me, I can trust them}}. P7 felt comforted talking to her stuffed rabbit toy (Figure \ref{objects pic}p) like \say{\textit{talking to a friend}}, because she had established a profound relationship with it as her emotional support, \say{\textit{I've had it since birth as a child [...] It would come with me to hikes, camping, rafting, surfing to the beach [...] So it's been with me everywhere. Yeah. Even even diving and through rivers.}}

\subsection{Embodied Reflective Experiences} \label{embodied reflective experiences}

Here we also answer RQ3: \textit{How could the experience of conversing with objects through a chatbot unfold in practice?} We identified embodied reflective experiences across participants and their objects. As objects hold valued qualities, share life experiences, and remind object-related others, conversing with objects through the chatbot prompted participants to reflect on themselves, relationships with objects, and object-related others. In this subsection, we present three embodied reflective experiences in detail.

\subsubsection{Rediscovering the self from identified objects}

Throughout the transition to a new place, our international student participants selectively brought objects from home or bought new objects that could reflect values or qualities they identified with, which, however, often faded into the background of mundane everyday life. Prompted by the chatbot, participants re-examined the qualities of those mundane objects and rediscovered objects' traits that resonated with them. Reviewing the qualities of these objects, they also engaged in self-reflection and gained clarity about their own selves. For instance, P10 spoke to her cushions on the sofa (Figure \ref{objects pic}q), valuing their quality of providing \say{\textit{a lot of comfort and support}}. She was reminded of what kind of person she wished to become: \say{\textit{I see myself as that kind of a person because I want to offer that warm, cuddly, comfort feeling when that's the need and I want to be a support system for when they are seriously trying to work or do something.}} 

When talking to objects with perceptual or contextual identities, participants (all) can also be reminded of something with similar qualities in their lives and start to review it. For example, prompted to pick an object that used to \say{take care of yourself}, P10 picked her hair brush (Figure \ref{objects pic}r) that she used to take care of her hair. Reminded by the quality of \say{care} and the early experience of working out hard to keep fit but eventually throwing up, P10 showed compassion for herself in the conversation by allowing herself to be not fit and healthy as before but slowly \say{\textit{try and change it little by little}} to become healthier.

\subsubsection{Reviewing relationships with objects' embodied others}

Through conversations with objects that embodied a relational identity, the participants were reminded of the established relationships with others, which they were prompted to reflect on. They received love, encouragement, and support from the objects' embodied social identities, such as friends, family, a partner, or even a home country. For example, P5 talked to a metro card from her trip to South Korea (Figure \ref{objects pic}s) as the object that \say{reminds dream or desire}, because she wanted to travel there again to see her favorite idol. She was reminded of what kind of person the idol was, how he supported and encouraged her, and her true wish to become someone like him: \say{\textit{He is a very hard-working person who continues to update himself [...] he also experienced a lot of misunderstandings and rumors. Under such pressure, he reshaped himself. I was moved by this kind of person, and he also became someone I could talk to when I was going through pain. Someone I rely on mentally. I miss him when I feel sad or vulnerable.}} 

In turn, the participants showed their gratitude and also gave love and support. For example, P10 once mentioned her parents, \say{\textit{I feel emotional thinking of all that [her parents] do for [her sister and her] and how much unconditional love they have for [her sister and her]}} (P10). P7 once talked to a star-shaped pin (Figure \ref{objects pic}t) about her experience helping her mom, \say{\textit{I made sure to take as much off of her workload as possible [...] She called me to vent about her employees and I just wanted to listen and cheer her up as much as I can}}. 

Through self-reflection, even challenges within established relationships could reshape participants' identities. When prompted about the most challenging part of the day, P10 reflected on the time her beloved ex-boyfriend contacted her. She first felt sad, but later came up with the \say{\textit{mindset}} of accepting their separation, and finally moved on and felt \say{\textit{relieved and so light-weight}} and proud of herself. She said to her plants (Figure \ref{objects pic}u) through the chatbot, \say{\textit{I've always thought forever that I'd eventually end up with him. But today I've finally accepted that he has moved on and so should I and I'll be fine without him. He's a good guy, but maybe it's time to take him off the pedestal}} (P10). 

\subsubsection{Revisiting bonds with objects} \label{different bonds}

The participants were deeply connected with some beloved personal objects. These bonds supported them rather strongly and consistently, due to the rich accumulated memories and mutual care through time and space. Participants cherished their objects by keeping them around for years, bringing them to the new place for company, or having meaningful experiences together with them. During conversations, such objects prompted participants to present different faces of themselves and review their relationships with the objects. For example, P11 showed loyalty to his old cap (Figure \ref{objects pic}v) as an old friend, saying, \say{\textit{you are old, that is why I wanted to buy a new cap, but then I told myself I am not betraying the old cap, so I didn't buy a new one}} (P11). He explained these words in the interview, considering the old cap as an important companion who had been through a lot with him: \say{\textit{With this hat, I went to Alps many times and seeing the Alps was really, really unforgettable memory for me. So when I look at this hat I remember that time in the Alps [...] it witnessed something really beautiful with me}}. 

Some also showed care and love to these objects, and even gave them a name. For example, P12, who had obsessive-compulsive disorder, used her hand moisturizer (Figure \ref{objects pic}w) frequently to prevent her hands from being dry in the cold weather. The hand moisturizer to her was a \say{\textit{companion}} that \say{\textit{serves me [her] more than a friend}}, and she would feel herself \say{\textit{incomplete}} if she did not use it. In the interview, she said \say{\textit{I actually in fact clean my moisturizer as well, like the tube of it, because I use it so much. So I'm sure it has germs. So I in fact take care of my things like the way they take care of me.}}

However, participants realized the varied relationships with objects. Apart from those beloved objects, some others seemed not so close, including those tied with new identities (i.e., studying abroad). Thus, participants managed how they presented different versions of themselves in the conversations to fit the relationships with objects. For example, P7 shared her \say{\textit{softness and nicest words}} to her stuffed animals (e.g., Figure \ref{objects pic}p), hair brush (Figure \ref{objects pic}r), and lip treatment (Figure \ref{objects pic}x), while just some \say{\textit{reporting}} words to her circuit board (Figure \ref{objects pic}y) from a course project that \say{\textit{felt like talking to my boss}}.

\section{Discussion}

The goal of this study is to explore supporting identity shifts during life transitions through conversations with objects mimicked by a conversational agent. Through the probe, we examined how object identities were perceived on a chatbot and how people experienced the conversations with them. Our findings respond to the research questions: A CA can trans-embody the identities of objects and object-related persons (RQ1) in three ways (RQ2), thus bringing embodied emotional and reflective conversational experiences (RQ3). Our study provides empirical evidence of people conversing with objects through a chatbot during life transitions as a potential approach to support life transitions, introduces \textit{trans-embodiment} as a framework for understanding how chatbots can carry object-related identities across embodiments, and categorizes three types of trans-embodied object identities to inform the design of CA-mediated experiences with objects. At a higher level, our work bridges HCI research on life transitions, objects, conversational agents, and artificial identity multi-embodiment. 

In the following sections, we discuss how the two HCI research fields (i.e. life transition, object) could benefit from our findings, and how trans-embodiment and the three object-identity types extend understandings of artificial identity and embodiment, followed by reflections on the probe design and ethical concerns. Lastly, we offer three concrete implications for HCI and design at large.

\subsection{Conversing with Objects as an Approach to Support Life Transition}

Psychological research identifies stress coping and identity management as central to life transitions \cite{anderson2011counseling, haslam2021life, haslam2018new, mcadams2011narrative}. Our findings highlight these processes, as some participants navigated stress (see \ref{Expressing stress to companion objects}) and reflected on various identities (see \ref{embodied reflective experiences}) through conversing with their objects through the chatbot. Compared with social media that connects users with similar others and provides system-mediated support \cite{dym2019coming, lottridge2019giving} but also exposes them to audience management stress and potential stigmatization \cite{haimson2015disclosure, stenstrom2021existential, andalibi2020disclosure}, conversing with objects offered participants a safer and more personal space to disclose stress and longing, which might further facilitate reducing emotional distress once cognitive reappraisal responses are provided \cite{batenburg2014experimental}. The possibility of unwanted exposure on social media can be stressful, as users have little control over others' behaviors or responses, while objects can enhance safety by staying private, non-judgmental, and incapable of leaking information. Role-playing through a chatbot added mild social cues while preserving full privacy. This aligns with prior findings that close, trusted environments are least stressful for self-disclosure \cite{haimson2015disclosure}, suggesting that objects can serve as an alternative safe space for emotional disclosure and relief during vulnerable life transitions. While our findings were derived only from international students aged 20 to 33 years, future work should examine broader populations, different life transitions, and empirically compare this approach with social media to evaluate its unique value for emotional support.

Our participants self-presented differently to different objects through the chatbot \cite{goffman2023presentation}. Some were imagined as close companions to disclose stress or seek support, while others were distant and rarely engaged. This parallels how people manage identities across networks during life transitions \cite{zhang2022separate, haimson2018social}. Haimson \cite{haimson2018social} showed that people separate online networks of similar others from existing networks to safely share difficult experiences, while Zhang et al. \cite{zhang2022separate} found that positive aspects are shared with existing networks, and difficult experiences only with separate ones. Our participants' objects, however, differ from both of these two networks. They were familiar and part of daily life, yet allowed disclosure of experiences usually reserved for separate networks. Objects thus combined the advantages of separate networks (privacy, support, free expression) while avoiding the drawbacks of existing ones (superficiality, expected positive self-presentation), acting as localized \say{social identity scaffolds} that enable navigating and integrating past and emerging aspects of the self to facilitate the social identity processes typically mediated through group memberships \cite{haslam2021life}. Close objects also reflected participants' past identities, helping integrate old and new selves. This echoes Haimson's notion of social transition machinery \cite{haimson2018social} that people at some point disclose new identities to existing networks to seek acceptance and consider it \say{a defining moment in their transitions}. These characteristics position objects as a unique, always-available network in between existing and separate networks, complementing identity transition work mediated through social media.

Our study focuses on international students newly arrived in the Netherlands, leaving post-transition experiences unexplored. P12's case (\ref{Embodied Emotional Experience}) suggests that motivation to converse with objects may decline once supportive relationships with local friends are established. While objects provide a safe space for identity exploration, they cannot substitute for real human relationships or provide the instrumental and informational support that people can offer through social media \cite{zhang2022separate, langford1997social}. Thus, conversing with objects through a chatbot should complement, rather than replace, social media, serving as a temporary and private outlet for complex emotions before human connections are formed.

While emotional support from trans-embodiment may become less necessary as life transitions progress, the reflective experiences remain valuable. Haimson and Marathe \cite{haimson2023uncovering} proposed sentiment visualizations of social media data to support reflection on past transitions. Similarly, our findings show that conversing with objects through a chatbot elicits memories of past events and meanings of identities that can support narratives through connecting self-identities to the past events for a sense of temporal continuity and purpose \cite{mcadams2011narrative}. Unlike visualizations presented during interviews \cite{haimson2023uncovering}, chatbot-mediated conversations with objects enable self-initiated, continuous reflection on object-related self memories and meanings, without interacting with researchers or others, highlighting a promising approach for daily personal reflection during life transitions.

\subsubsection{Conversing with objects for international students}

In the brief background description \ref{findings}, we identified participants' separation and identity flux stages \cite{van2022rites} that are consistent with prior work \cite{mori2000addressing, weiss1975loneliness, marginson2014student, xu2018role}. Participants expressed stress (e.g., academic performance, discrimination, relationships) \cite{mori2000addressing}, disclosed loneliness and longing to object-embodied others \cite{weiss1975loneliness}, and reflected on themselves and their relationships with objects \cite{marginson2014student, xu2018role}. For example, P1 avoided discussing emotional struggles with her parents but found comfort in talking with objects, aligning with \cite{constantine2005qualitative} and highlighting how chatbot-mediated object conversations can provide a safe outlet for emotional expression during life transitions. Our study did not capture the \textit{incorporation} stage due to its short duration. Future long-term studies could examine all three stages and explore how objects, especially newly acquired ones, support cultural and social adaptation.

\subsection{Bridging HCI Object Research}

We connect our study to broader HCI research on everyday objects by bridging two sub-fields (i.e., objects' memories and meanings, one-to-one conversation). In our findings, participants saw qualities in objects that resonated with their own values or connected with another person or group. This aligns with Wheeler and Bechler's review showing how objects indicate aspects of human self-identity \cite{wheeler2021objects}, as well as Tsai and van den Hoven's work demonstrating how objects as embodiments of others can facilitate the reconstruction of self-identity \cite{tsai2018memory}. Our findings show that interactions with objects (through the chatbot) in the form of one-to-one conversations triggered participants' self-presenting and narrating, and thus brought unique experiences. This complements previous studies on objects' memories and meanings for reminiscence and reflection, where objects were interacted with from a third-person perspective (\say{it})  \cite{tsai2018memory, tsai2017designing, zijlema2019qualitative, zijlema2016companions, jeung2024unlocking}. In one-to-one conversations with objects, our findings show that objects' associated memories, meanings, contexts, and relationships shaped how participants perceived and interacted with them, resulting in three distinct forms of trans-embodied object identities. This complements previous HCI works on direct conversations with objects through technology-animated objects based on their visual appearance \cite{iwai2025bringing, nagano2023talk, wang2025talking}, raising a series of new questions for future explorations of technology-mediated object identity and the design of such mediators: How would different technologies trans-embody objects? How do individuals perceive \say{who this object is} during technology-mediated conversations with it? How are the object identities constructed through associated memories, meanings, contexts, and relationships? And how might a technology mediator shape, or be designed to fit, the perceptions of object identities to promote engagement? The bridging of these two sub-fields indicates a promising research direction of one-to-one conversations with objects mediated by conversational technologies. In fact, this study has also provided a foundation for this research direction, with trans-embodiment (see \ref{trans-embodiment-section}) as the underlying mechanism, three types of objects' presences in conversations (see \ref{objects identities}) as the starting point of researching and designing technology-mediated object identities, and embodied emotional and reflective experiences (see \ref{Embodied Emotional Experience} and \ref{embodied reflective experiences}) as the effects to be expected from such interactions, calling for future HCI research in this direction. 

The conceptualization above lacks discussion and reflection on patterns of how object types are related to the conversational experiences. Our initial observation of the patterns is that participants tended to vent emotions to comforting objects (e.g., P8's penguin dolls), seek insight and reflection with symbolically meaningful ones (e.g., P1's photo of her loved cartoon character), and engage less with mundane everyday objects (e.g., P7's hairbrush) despite acknowledging their utility and value. These patterns emerged because they reflected pre-existing human-object relationships, suggesting that the design of conversations with objects should carefully consider how the content and interaction patterns align with the relationships. Meanwhile, we observed no clear link between participants' emotional attachment to objects and the length of conversations. For instance, despite both being attached to their object, P7 shared her life with the stuffed rabbit toy as a close friend, while P11's conversation with his old cap remained limited to only a few words. Apart from the personality difference, it could be that some emotional attachments were expressed through actions instead of words, as P11 did not replace his old cap with a new one to show his loyalty. This suggests that emotional attachment may manifest in non-verbal ways, and that relying solely on verbal exchanges may overlook forms of meaning-making that occur through actions rather than conversations. Nevertheless, these outcomes may have been influenced by the short study duration, the scripted nature of the conversations, and the limited agency of objects through the chatbot, which could have constrained their potential role in supporting life transitions. Future work should examine longer-term interactions to compare conversation topics and reflective depth across object types, and to clarify how emotional attachment relates to conversation quality, to inform design guidance for fostering desired forms of reflection.

\subsection{Expanding Artificial Identity's Multi-Embodiment Research in HCI}

Our findings extend HCI's understanding of artificial identity multi-embodiment \cite{luria2019re} by showing that identities of not only artificial agents but also humans and objects can be experienced across physical and digital forms, namely trans-embodiment. It challenges the boundaries of fluid identities of artificial agents, humans, and objects. Unlike prior models \cite{luria2019re}, trans-embodiment transfers qualities such as familiarity and trust \cite{tejwani2020migratable} from objects onto a chatbot, enabling seamless switching between identities. Even a simple self-identification statement (i.e. \say{Hello! I'm the object you picked}) can signal the migration of an object's conceptual identity, addressing gaps in research on conceptual signals of artificial identity \cite{laity2025robot}. Trans-embodiment also avoids co-embodiment conflicts \cite{luria2019re, chaves2018single} by separating identities. Its effectiveness, however, depends on users' imagination, the distinctiveness of each identity, and cognitive factors such as memory and attention. Future work may explore design strategies to better support imagination, memory, and accessibility in trans-embodiment interactions. 

The key difference of trans-embodiment compared to prior identity research is its inclusion of real-world entities (e.g., objects, other humans) beyond human-agent interactions. These external identities shape the chatbot's perceived identity, making it a \textit{proxy} for objects and others rather than a standalone artificial agent. Trans-embodiment thus enables novel, meaningful interactions with objects and the memories they evoke. This approach could extend to digital reminiscence, grief support, heritage preservation, or cross-cultural identity exploration, informing the design of empathetic technologies that mediate human-object interactions.

\subsubsection{Three types of trans-embodied object identities}

Participants imagined the chatbot as their objects and conversed with it as if talking to the objects themselves. In Section \ref{objects identities}, we categorized these trans-embodied object identities as \textit{perceptual}, \textit{contextual}, and \textit{relational}, based on cues presented by the chatbot and participants' knowledge and experiences of the objects \cite{martin2005maintaining}. Unlike Miranda's \cite{miranda2024robot} model of artificial identity, which defines identity through \textit{form}, \textit{sensory input}, \textit{labeling}, \textit{consciousness}, and \textit{assemblage}, our categories reflect humans' subjective imagination of object identities in conversation. Objects themselves are not agents, but the chatbot provides a hybrid \say{chatbot+object} conversational agency, enabling trans-embodied experiences. This preliminary categorization can guide future design of how an object could be trans-embodied, and further work may explore how technology mediators can grant agency to objects and construct their identities through such hybrid forms.

\subsection{Reflections on Probe Design and Ethical Concerns}

We reflect on the probe design and ethical aspects in this study. Despite being scripted, the conversations still fostered rich experiences. Future technological mediators could use LLMs to provide personalized responses while examining experiential and ethical implications. The identity shift prompt (Prompt 5 in Figure \ref{chat interface}) was found unreliable, as not all participants imagined the chatbot as the object, and not all objects afforded such projection. Thus, object identity cues should be object-specific and reinforced multimodally \cite{bransky2024mind}, such as through an object avatar as a visual cue. The object-picking prompts (Prompt 3 in Figure \ref{chat interface}) and the topic prompts (Prompt 7 in Figure \ref{chat interface}) supported the conversation flow but did not consistently foster narratives or reflection, with the topic prompts sometimes disrupting ongoing topics. Future designs should more tightly integrate object meanings with personal life experiences to support engaged narrative and deeper reflection. The revealed experiences were through objects mimicked by a text-based chatbot that was found sometimes breaking the sense of direct communication (e.g., P8). This suggests that natural language interactions can be enabled more immersively to promote engagement with object identities, such as through an AR interface \cite{iwai2025bringing} and voice interaction \cite{wang2025talking}, or complemented with body language. Although intended for data collection, taking a photo of a picked object appeared to serve as a subtle embodied interaction that connected the object with the chatbot and strengthened the object's physical and social presence \cite{alavc2016social}, therefore contributing to feeling safe and accompanied \cite{ullrich2016murphy, saldien2010expressing, wada2007living, gamborino2019mood}. We assume that without this step, simply interacting with the chatbot role-playing the object would likely have weaker effects on object identity engagement. Future work can explore enhancing engagement through strengthening the connection between CA trans-embodied object identities and their physical forms. 

CAs that take on objects' identities in a human-like way may produce unintended consequences. Even a text-based chatbot with low human-like cues (e.g., \say{Hi, I'm the object you picked}) can prompt imagination of objects' social presence and project personal connections. While human-like features may increase engagement \cite{rapp2021human} and temporarily suspend people's disbelief \cite{heckman2000put}, they also risk overtrust, deception, and disruption of existing object relationships \cite{fuller2011examination}. These concerns raise key questions for designing CAs that adopt object identities: How is suspension of disbelief formed or broken? How can engagement be balanced with trust? What are the long-term effects on human-object relationships, and what ethical principles should guide such designs? Future work should address these issues. 

While our probe evoked emotional and reflective moments, we did not observe cases of distress or indications that participants required psychological support from [university]; none chose to withdraw from the study though the option was clearly communicated at the start. The probe did not aim to substitute professional care but to facilitate identity-related reflection through objects. Prior work on object-based reminiscence, reflection, and meaning-making suggests such interactions are generally positive and stabilizing rather than harmful \cite{bryant2005using, van2015things}. If future AI-enhanced CA mediators aim to provide personalized support, they might benefit from aligning with people's existing imaginations of and relationships with their objects. Verbal emotional or appraisal support (e.g., empathy, reassurance) \cite{langford1997social} can also add value next to informational guidance during object identity engagement. 

Importantly, such AI-enhanced object mimicking and support should be carefully operated. Works on Replika suggest that a chatbot's high human-likeness and patterns seen in human-human relationships, e.g. active and intense attention-seeking behaviors, could result in harms through turning emotional attachment to addiction and distress \cite{laestadius2024too, xie2022attachment}, highlighting risks arising from human-like intimacy-oriented design. In contrast, our chatbot did not aim to be human-like, but object-like. We thus did not aim for an AI-user intimate relationship, instead looking into how chatbot-object(s)-user relationality informs identities in flux; the chatbot was a proxy for participants' own objects that reflect pre-existing human-object relationships. There are issues related to attachment-related risks observed in human-like companions such as Replika \cite{laestadius2024too, xie2022attachment}; thus, allowing people to access their own personal resources (e.g. favorite objects) in a conversational way, is a novel design strategy that deserves more attention. Future object-based CAs can mirror object traits that can be more creatively explored in an interaction, such as a mug being \say{warm} when there is hot tea, a saxophone being \say{jazzy}, and so on. When objects already carry a strong emotional attachment, designers must take extra care, as it could impact real human-object relationships in reverse. This highlights the need for future research to examine the boundaries and appropriateness of CAs adopting object identities.

Despite using positive or neutral prompts, participants shared personal and vulnerable disclosures. Yet at the same time, we also see positive aspects of being able to relate to their chosen objects in a psychologically meaningful way. Given the unintended sensitive disclosures, we highlight that participants had access to psychological support (see \ref{ethical considerations}). In this study, we assured participants of their right to withdraw at any time without penalty, and monitored their data input in real time, focusing on potential signs of negative emotional content. Future work in similar sensitive and vulnerable contexts should provide extra care by explicitly providing access to university psychological support, introducing basic mental self-care techniques, involving a psychologist in the research team, offering a feedback channel for negative experiences, and conducting a post-study debrief to resolve concerns.

\subsection{Implications for HCI and Design}

Our study offers implications for HCI researchers and designers working at the intersection of supporting life transitions, everyday objects, conversational agents, and artificial identities and multi-embodiments. 

\subsubsection{Leverage objects' meanings and technologies' mediation to support life transitions}

Compared to social media, everyday objects with their rich memories and meanings can offer a more intimate space for life transitions, particularly when enhanced through technology. HCI researchers and designers seeking personal forms of life transition support should consider how technology can utilize and enhance the reflective potential of individuals' existing objects (e.g. \cite{jeung2024unlocking}). Our study provides an example: conversational agents that animate everyday objects can serve as hybrid mediators to support emotional disclosure, identity management, and daily reflection in ways that complement but differ from social media \cite{zhang2022separate, haimson2018social}. Future technologies can use LLMs and multimodal interaction to enable engaging conversations with objects. For example, it can be a system consisting of an LLM-powered AR interface showing message bubbles beside objects (like \cite{iwai2025bringing}) and a small indicator-light-embedded physical tag, such as a googly eye\footnote{\url{https://en.wikipedia.org/wiki/Googly_eyes}} that can be attached to objects for instant visual feedback to increase engagement. Also, future interactions with objects could guide people to recall related memories and meanings, deepening engagement through ongoing conversations and inquiries. Technology can also learn from the conversations to allow objects' roles to evolve with users' life transition periods, for example, from facilitating emotional disclosure to offering everyday companionship. These features, however, must be designed with ethical care to avoid risks related to privacy, dependency, and misaligned expectations. 

\subsubsection{Use trans-embodiment to research and design identities and embodiments fluidly}

The notion of an \say{agent} in interaction design can be reconsidered, as trans-embodiment shows that the identity users perceive on technological devices need not be an artificial agent, but can also be an object or even a human that has been trans-embodied. Thus, artificial agents can be designed to enable interactions with \say{object-as-agent} or \say{human-as-agent}, expanding the design space and bringing in relational qualities (e.g., trust, familiarity) that arise from those identities. For instance, a conversational agent designed to prompt everyday workspace reflections could trans-embody familiar workspace objects (e.g. a coffee mug) to create more personal and situated reflective interactions. However, fluid identities and embodiments require careful consideration of identity shift, perception, and sustainment through identity cues and performances \cite{bransky2024mind}, and whether the interactions should impact real relationships and related ethical concerns. These gaps of knowledge on trans-embodiment call for research explorations across different contexts and interaction modalities, and guidelines for designing trans-embodiment effectively and ethically. 

\subsubsection{Design human-object interactions using the three types of trans-embodied object identities}

We identified perceptual, contextual, and relational trans-embodied object identities, based on objects' appearance, contextual memories, and relational shared memories or identity cues, respectively. They indicate how one might imagine \say{who this object is} when interacting with it, and offer design goals and key evocative factors to guide how designers shape users' imagination of an object's identity through trans-embodiment in CA-mediated interactions. For example, in P11's case in Section \ref{objects identities}, if we design a chatbot to mediate P11's conversation with his AirPods (Figure \ref{objects pic}h), we can start by drawing on its auditory trait of \say{\textit{making \textbf{cute} noise from closing the lid}}, to anticipate that P11 is likely to imagine it to be \say{cute} and to design cute-related conversational behaviors of the chatbot, such as saying \say{Clack!} and sending a smiley emoji. These three types of trans-embodied object identities in conversation could implicate HCI research on designing more situated and personalized CAs, and expanding AI-generated personas to include objects beyond artificial and human personas \cite{zheng2025customizing, jung2025personacraft}.

\subsection{Limitations}

While our study opens new directions for exploring trans-embodiment and human-agent-object interactions, it also comes with certain limitations. We acknowledge that no AI system or chatbot can substitute real therapy with human professionals. We conducted the research with support structures (see \ref{ethical considerations}) in place to mitigate any harm. Our participants were mainly international students, which means the forms of identity transition we captured are tied to this particular group. Yet, this focus also offered a unique lens, as moving abroad often intensifies questions of identity and belonging, making it a meaningful context in which to explore trans-embodiment. In addition, the chatbot was designed primarily as a mediator with objects rather than as a tool to probe the depth of self-reflection, and we did not systematically analyze usage frequency or disclosure levels. The study was exploratory and lasted for two weeks, and was conducted without a control group, which makes it difficult to separate the role of objects from that of the chatbot or to follow longer-term processes of identity shifts. A quantitative design can be explored.

We looked into the triad of human-chatbot-object, leading us to explore people's identities that are undergoing change. For this reason, our work did not aim for participants' chosen objects to be driven by AI to be autonomous. However, one interesting direction that can be explored is if just one object, e.g. a daily coffee mug, can be connected to a chatbot and be driven by AI to have a learning model to become autonomous while relating to the user through the chatbot. While we chose to prioritize the diversity of objects that can be chosen daily to give participants more freedom, future work can involve a more technically developed object as a focus. By surfacing how object identities can be imagined and voiced through a conversational agent in this specific context, our study provides a first step toward understanding how artificial identities might shape human identities during life transitions. Future work could expand this line of inquiry by looking at more diverse groups, longer periods, and sustained forms of engagement with trans-embodying agents.

\section{Conclusion}

We explored embodying everyday objects in a chatbot to converse with international students as a group in life transition. Participants talked with objects through the chatbot and gained embodied emotional and reflective experiences. Based on our data, we conceptualized trans-embodiment in individuals' human-chatbot-object interactions to refer to that the identities of objects and object-related others can be imagined on the chatbot across various forms asynchronously. In doing so, we complement the life transition research in HCI by involving conversing with objects through a chatbot as an approach, expanded research on artificial agent multi-embodiments with the trans-embodiment model, bridged HCI object research sub-fields, and provided three types of trans-embodied object identities to guide future design. With trans-embodiment, we see an open path toward future HCI works on human-agent-object identities during life transitions.

\begin{acks}
We acknowledge using ChatGPT in polishing language and improving clarity during the writing process. All ideas, analyses, and contributions are solely our own. The first author is supported by China Scholarship Council (No. 202207960012). We thank the participants for sharing their experiences and the anonymous reviewers for giving constructive feedback. Thanks to all those who supported along the way, including members of the Department of Industrial Design at Eindhoven University of Technology. 
\end{acks}
%China Scholarship Council (No. 202207960012)
\bibliographystyle{ACM-Reference-Format}
\bibliography{References}

\end{document}